\pdfoutput=1

\documentclass[final,1p,times]{elsarticle} 

\usepackage{graphicx}
\usepackage{amssymb} 
\usepackage{amsthm} 
\usepackage{lineno}

\newcommand{\commentout}[1]{{}}

\journal{Nuclear Physics A} 

\begin{document}

\begin{frontmatter} 

\title{Hydrodynamic fluctuations and two-point correlations}


\author{Todd Springer}
\author{and Mikhail Stephanov}
\address{Department of Physics, University of Illinois at Chicago}

\begin{abstract} 
  We examine correlations of energy density induced by initial state
  fluctuations, which are localized in both transverse and
  longitudinal extent.  The hotspots are evolved according to
  hydrodynamics in a background which includes radial flow.  Two-point
  energy density correlations from these hotspots are computed as a
  function of the difference in azimuthal angle and rapidity.  Such
  localized perturbations occur naturally in the theory of
  hydrodynamic fluctuations and may provide insight into some features
  of the two-particle correlation data from RHIC and the LHC.
\end{abstract} 

\end{frontmatter} 


\section{Introduction}
Two-particle correlation measurements in heavy ion collisions
\cite{STARTwoParticle,ALICETwoParticle,CMSTwoParticle,ATLASTwoParticle}
contain many qualitatively different features than those in pp
collisions; the hydrodynamic nature of the quark-gluon plasma has a
significant effect on correlations.  We are interested in the
impact of \emph{local} fluctuations on two-particle correlation data
from RHIC and LHC.  In contrast to typical initial state fluctuations,
which are assumed to be nearly boost invariant, here the fluctuations
are localized in both transverse \emph{and} longitudinal extent.  In
\cite{Staig:2011wj} it was shown that initial state hotspots placed
at a particular position in the transverse plane could reproduce many
of the features of the two-particle azimuthal correlations.  Our
approach follows this work closely, but allows us to also examine the
rapidity extent of the correlations.

Point-like fluctuations occur naturally due to thermal noise in
hydrodynamics.  In \cite{Kapusta:2011gt} it was shown that local
hydrodynamic fluctuations occurring throughout the evolution of the
system could induce long range rapidity correlations.  These
proceedings are a first step toward the extension of these results to
include the angular dependence of the correlation functions.  

\section{Background flow}
In the same manner as \cite{Staig:2011wj}, we consider an exact
solution to the equations of hydrodynamics which incorporates radial
flow.  The solution we use is the Gubser flow
\cite{Gubser:2010ze,Gubser:2010ui}.  The equations of hydrodynamics
are solved exactly for a fluid which is at rest in the coordinate system
\begin{eqnarray}
  ds^2 = \tau(\rho,\theta) \left[-d \rho^2 + \cosh^2 \rho \left(d \theta^2 + \sin^2 \theta d\phi^2 \right) + d \xi^2 \right].
\end{eqnarray}
The angle $\phi$ is the usual azimuthal angle around the beam line, and
$\xi = \tanh^{-1}(z/t)$ is the space-time rapidity.  The ``conformal
coordinates'' $\rho, \theta$ are related to the usual proper time
$\tau = \sqrt{t^2 - z^2}$, and radial distance from the beam line $r =
\sqrt{x^2 + y^2}$ by the coordinate transformations
\begin{eqnarray}
  \sinh \rho = - \frac{1-(q \tau)^2 + (q r)^2}{2 q \tau} \hspace{15mm} \tan \theta = \frac{2 q r}{1+ (q \tau)^2 - (q r)^2}.
\end{eqnarray}
These coordinate transformations contain a free parameter $q$ which
has dimensions of energy; it allows one to specify a scale which is
related to the transverse size of the system $R \sim q^{-1}$.  In the
limit $q \to 0$, the system has infinite transverse extent and the
Bjorken solution is recovered.  The radial velocity of the fluid is given by
\begin{eqnarray}
  v_r = \frac{2 q^2 r \tau}{1 + (q \tau)^2 + (q r)^2}.
\end{eqnarray}

The Gubser flow is an idealization of the quark-gluon plasma formed in
a heavy ion collision.  In particular, the solution requires azimuthal
symmetry around the beam line and conformal symmetry.  The former is
restricts us to central collisions only; the latter is only
approximately true for $T \gg T_c$, where $T_c$ is the crossover
temperature.

\section{Perturbations of the flow}
We add small perturbations of energy density and fluid velocity to the background flow,
\begin{eqnarray}
  \varepsilon(\rho,\theta,\phi,\xi) = \frac{\hat{\varepsilon}_0(\rho)}{\tau^4} \left[1+ \delta(\rho,\theta,\phi,\xi) \right], \\
  u^\mu(\rho,\theta,\phi,\xi) = \frac{1}{\tau} \left(1, u^i(\rho,\theta,\phi,\xi) \right).
\end{eqnarray}
The background energy density $\hat{\varepsilon}_0$ is found by solving
the equations of hydrodynamics in the absence of the perturbations.

\begin{figure}[htbp]
\begin{center}
\includegraphics[width=0.95\textwidth, trim = 0mm 15mm 0mm 16mm, clip=true]{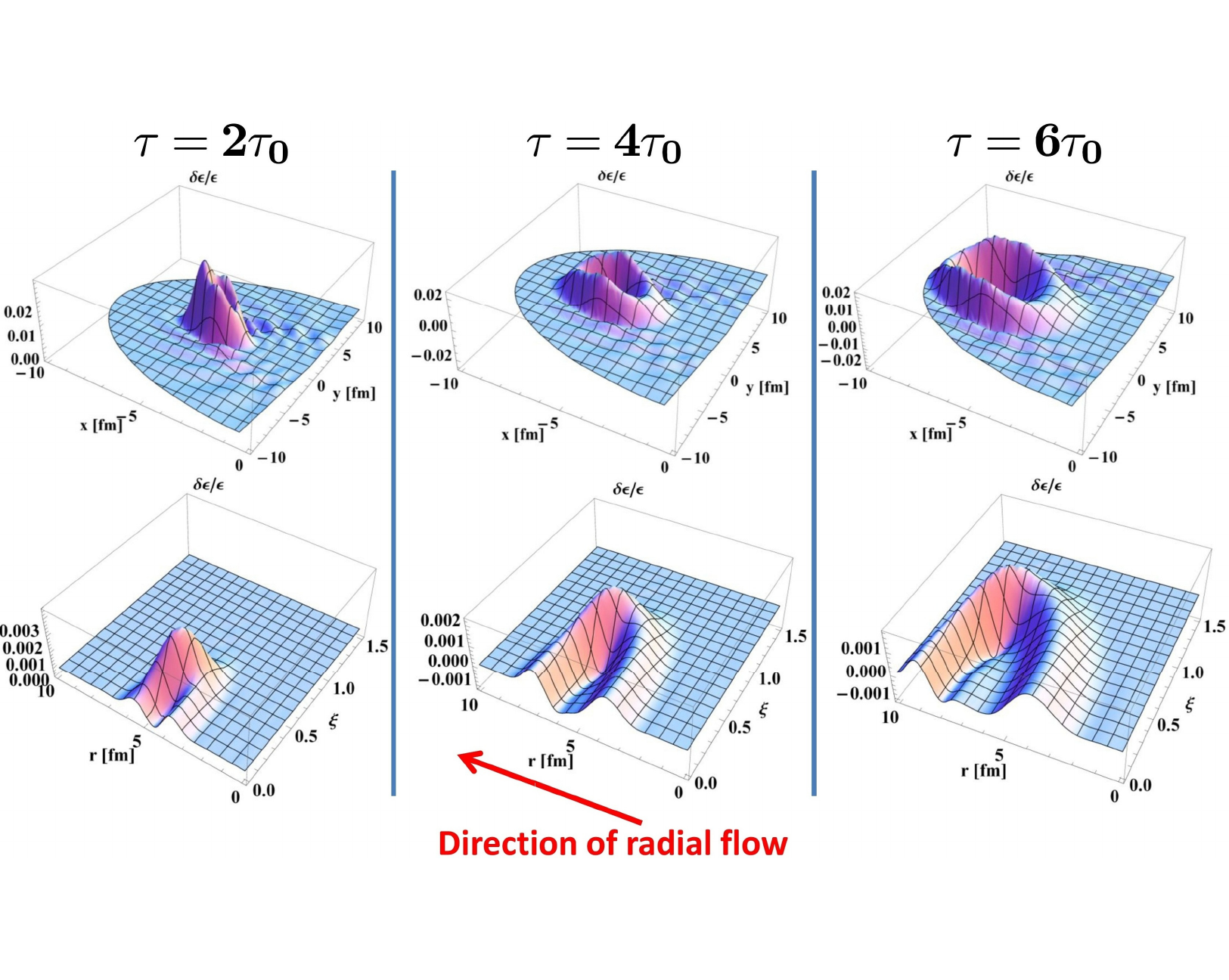}
\end{center}
\caption{Plots of the evolution of a hotspot.  The columns
  show the profile of energy density at three different times.  The
  upper row shows the evolution in the transverse plane (after an
  integration over rapidity), while the lower row shows the evolution
  in the $r-\xi$ plane (after an integration over azimuthal
  angle).  }
\label{EvolutionFigure}
\end{figure}

To reduce the linearized hydro equations to a set of ordinary differential
equations, we employ a spherical harmonic expansion in the
$\theta,\phi$ coordinates, and a Fourier transform in the longitudinal
direction\footnote{Note that we have reduced $u^\theta$ and $u^\phi$
  to the gradient of a single scalar function.  In fact, there is one
  additional divergenceless vector component to the velocity, but it
  does not couple to the equations for the energy density.  In this
  work, we will present results for the correlations of energy density, so
  we do not need to consider the vector piece.},
\begin{eqnarray}
  \delta(\rho,\theta,\phi,\xi) &=& \int \frac{dk}{(2\pi)}\sum_{lm} \delta^l(\rho) Y_{lm}(\theta,\phi) e^{i k \xi}, \\
  u^\theta(\rho,\theta,\phi,\xi) &=& \partial_\theta \int \frac{dk}{(2\pi)}\sum_{lm} u_S^l(\rho) Y_{lm}(\theta,\phi) e^{i k \xi}, \\
  u^\phi(\rho,\theta,\phi,\xi) &=& \frac{1}{\sin^2 \theta}\partial_\phi \int \frac{dk}{(2\pi)}\sum_{lm} u_S^l(\rho) Y_{lm}(\theta,\phi) e^{i k \xi}, \\
  u^\xi(\rho,\theta, \phi,\xi)  &=& \int \frac{dk}{(2\pi)}\sum_{lm} \nu_\xi^l(\rho) Y_{lm}(\theta,\phi) e^{i k \xi}.
\end{eqnarray}
There are then three coupled equations for the perturbations
$\delta^l, u_S^l, \nu_\xi^l$, which are given in \cite{Gubser:2010ui}.
\commentout{
\begin{eqnarray}
  &&\partial_\rho \delta^l - \frac{4l(l+1)}{3} u_S^l \rm{sech^2}(\rho) + \frac{4 i k}{3} \nu_\xi^l = 0 \\
  &&\partial_\rho u_S^l + \frac{1}{4} \delta^l - \frac{2}{3} u_S^l \tanh \rho  = 0 \\
  &&\partial_\rho \nu_S^l + \frac{ik}{4} \delta^l - \frac{2}{3} \nu_\xi^l \tanh \rho  = 0
\end{eqnarray}
} 
When viscosity is neglected, the equations can be solved
analytically.  We assume an initial condition at conformal time
$\rho = \rho_0$ such that the energy density is initially localized,
\begin{eqnarray}
  \delta(\rho_0, \theta,\phi,\xi) &=& A \exp \left\{- \frac{\theta^2 + \theta_0^2 - 2 \theta \theta_0 \cos(\phi - \phi_0) + \xi^2}{2 \Sigma^2} \right\},
\end{eqnarray}
and all fluid velocities vanish.  The solutions are then transformed
back to the more familiar $\tau,r$ coordinates.  We plot the time
evolution of an example initial state hotspot in Fig. \ref{EvolutionFigure}.
\section{Correlations}
The solution to the differential equation for the over-density
$\delta$ at a final time $\tau_f$ is a function of three coordinates
$r,\phi,\xi$.  Usually, to translate this coordinate space result into
momentum space (relevant for comparison with experiment), a
Cooper-Frye freeze-out is performed which involves integration over
all coordinates.  We will defer a full freeze out analysis to future
work and here simply integrate only over the radial distance at
constant $\tau_f$.  Such an integration would arise from the
Cooper-Frye prescription assuming that matter at different radial
positions freezes out at a fixed proper time $\tau_f$.  As such, we define
\begin{eqnarray}
  \bar{\delta}(\phi, \xi) \equiv \int_0^\infty r dr\, \delta(\tau_f,r,\phi,\xi).
\end{eqnarray}

\begin{figure}[htbp]
\begin{center}
\includegraphics[width=0.6\textwidth]{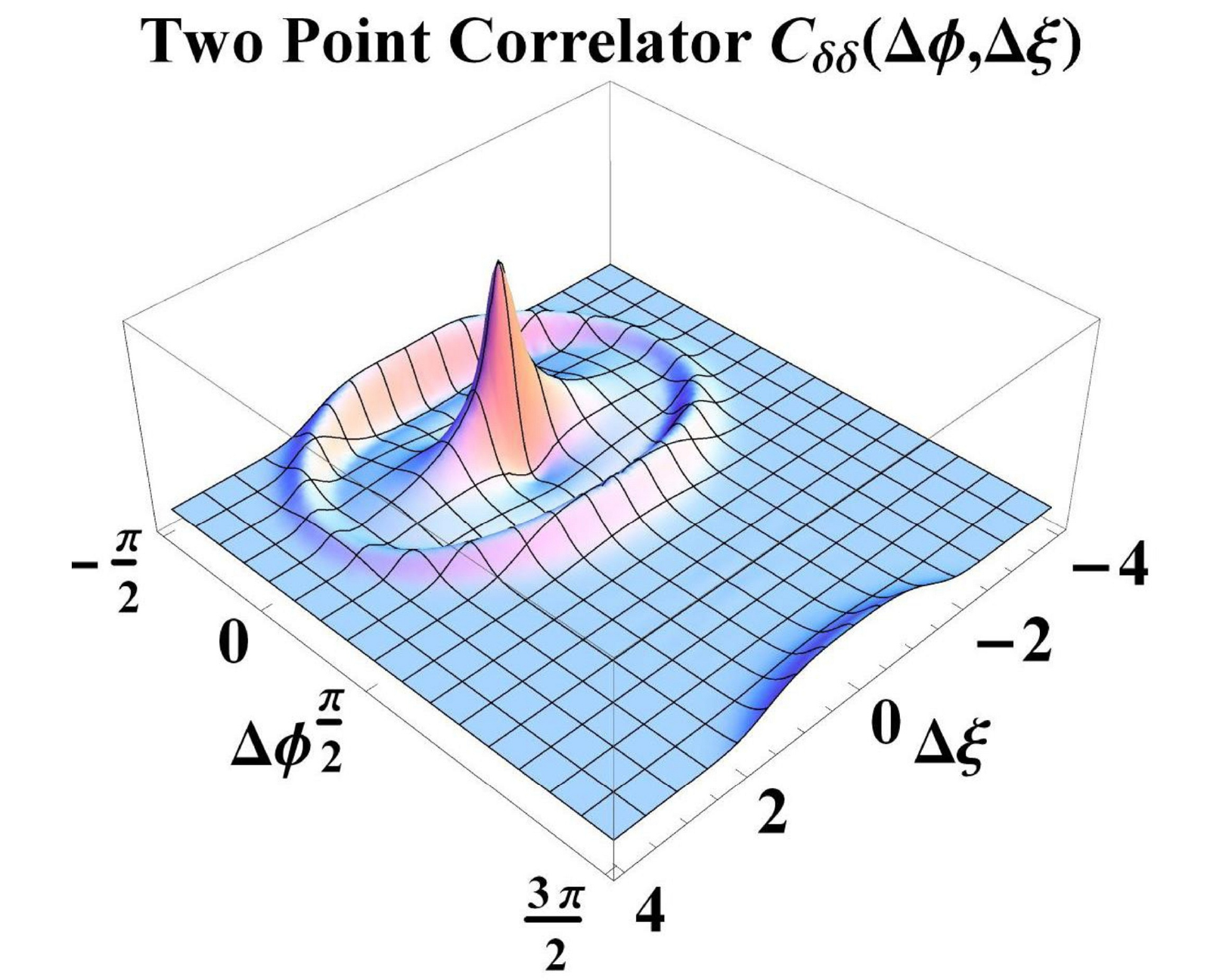}
\end{center}
\caption{Plot of the correlation function of energy density as a function of the (coordinate space) angular difference and spacetime rapidity difference.  To make this plot, we summed harmonics up to $l = 25$.  The parameters used were $\tau_0= 0.5\,{\rm fm}/c$, $\tau_f = 10\, \rm{fm}/c$, $q^{-1} = 4.3 \, \rm{fm}$, $\Sigma = 0.1$, $A = 1/2$.  Following \cite{Staig:2011wj}, the initial radial position of the hotspot was $r_0 = 4.1 \, \rm{fm}$. }
\label{CorrelationFigure}
\end{figure}

We compute the correlation function by multiplying two such solutions
and integrating over the shift of azimuthal angle and rapidity.  In
this way, we are integrating over all hotspots placed at varying
initial angles and rapidities.  As in \cite{Staig:2011wj}, for
simplicity, the initial radial position of the hotspot, $r_0$, is fixed
and not integrated.  The correlator,
\begin{eqnarray}
  C_{\delta \delta}(\tau_f, \Delta\phi, \Delta \xi) \equiv \int_0^{2\pi} d \chi \int_{-\infty}^{\infty} d \eta\, \bar{\delta}(\tau_f,\phi_1 - \chi, \xi_1 -\eta) \bar{\delta}(\tau_f,\phi_2 - \chi, \xi_2 -\eta).
\end{eqnarray}
is thus a function of the angular difference $\Delta \phi = \phi_1 -
\phi_2$ and rapidity difference $\Delta \xi = \xi_1 - \xi_2$.  The results are shown in Fig. \ref{CorrelationFigure}.

\section{Summary}
We have examined the energy correlations induced by an initial hotspot
localized in both transverse and longitudinal extent.  Because the
hotspot propagates according to hydrodynamics, a sound front is
evident in Fig. \ref{CorrelationFigure}.  In addition to the sound
front, we notice a peak at zero separation which falls off more slowly in the rapidity direction than in the angular direction.  These properties may have some
implications for near side peak observed in two-particle correlation
measurements at RHIC and LHC.

Rather than introducing the initial source by hand, this approach
should be extended to include stochastic ``noise'' in order to make this 
approach consistent with the theory of hydrodynamic fluctuations
\cite{Kapusta:2011gt}.  It is also necessary to perform a proper
freeze-out to translate the correlations in coordinate space to those
in momentum space (which are observed in experiment).  These
investigations are currently underway.

\section*{References}
\bibliographystyle{c:/Users/Todd/Documents/Physics/Hydro/Bibliography_Files/h-elsevier3}
\bibliography{c:/Users/Todd/Documents/Physics/Hydro/Bibliography_Files/hydro}


\end{document}